\documentclass[aps,final,notitlepage,oneside,twocolumn,nobibnotes,nofootinbib,
superscriptaddress,noshowpacs,centertags]{revtex4-1}

\usepackage[cp1251]{inputenc}
\usepackage[english]{babel}
\usepackage{graphicx}
\usepackage{latexsym}
\usepackage{amssymb}
\usepackage{amsmath}
\usepackage{float}

\begin{document}

\title{Gravitational waves from primordial black holes collisions in binary systems}
\author{Yu.\;N.\;Eroshenko}\thanks{e-mail: eroshenko@inr.ac.ru}
\affiliation{Institute for Nuclear Research of the Russian Academy of Sciences
60th October Anniversary Prospect 7a, 117312 Moscow, Russia}

\date{\today}

\begin{abstract}
It was shown by (Nakamura et al. 1997), (Ioka et al. 1998), and (Sasaki et al. 2016) that primordial black holes (PBHs) binaries can form effectively at the cosmological stage of radiation dominance, and the merge of the PBHs in pairs can explain the gravitational wave burst GW150914. In this paper, the model is re-examined by considering the effect of inflationary dark matter density perturbations which produce additional tidal forces. 
As a result, the merge rate of PBHs binaries and the corresponding rate of the gravitational bursts are suppressed by the factor $\sim1.5-2$ in comparison with previous calculations. This rate matches the LIGO data if the PBHs constitute the $f\sim5\times10^{-4}-5\times10^{-3}$ fraction of dark matter. 
\end{abstract}

\maketitle 

%\tableofcontents

%\bigskip

%%%%%%%%%%%%

\section{Introduction}

Although gravitational waves were predicted by Albert Einstein in 1916, to date their existence was shown only indirectly through the 
orbital changes of binary pulsar PSR~B1913+16. The first direct detection was done 
September 14, 2015 by two laser interferometers LIGO \cite{GW150914}. The form of the GW150914 signal corresponds to the 
general relativity prediction for the merge of two black holes with masses $36M_\odot$ and $29M_\odot$, and the statistical significance of the registration is $5.1\sigma$. This result demonstrates the existence black holes in the binary systems. Such black holes can born at the massive supernova explosions as a result of the standard stellar evolution \cite{Grietal01,BogLipTut07}. In particular, the low-metallicity environment boosts the massive double black holes formation \cite{chris1,chris2,LigoVirgo}. It is possible, however, that the black holes were born in the collapse of massive pre-galactic stars, or they were formed by merge of smaller black holes in the dense star clusters.

The merge of primordial black holes (PBH) in a pairs provides the alternative explanation of GW150914 \cite{Naketal97,Ioketal98,Sasetal16}. The possibility of PBH formation was predicted in general by \cite{ZelNov67,Haw71}, and there are many particular scenarios of their formation \cite{Car75,GarLinWan96,Jed77,KhlPol0,BerKuzTka83,DolSil93}. For concreteness we consider in this paper the formation of PBHs from curvature perturbation as in \cite{Car75}. 

The PBH pair
can form occasionally if two PBH born sufficiently close to each other \cite{Naketal97}. The motion 
of the binary PBH is influenced by neighbouring third PBH, therefore the PBHs do not move exactly towards each other but experience
tangential displacement under the influence of the tidal gravitational forces from the third PBH. The \cite{Naketal97,Ioketal98} discussed the MACHOs with masses $\sim0.1-1M_{\odot}$, while \cite{Sasetal16} explained the signal GW150914 as the merge of PBHs with masses $\sim30M_\odot$.  In addition, the possibility of the PBHs clusters formation in the early universe was proposed by \cite{RubSakKhl01}, and the PBHs collisions in such clusters can produce the bursts of gravitational waves \cite{DokEroRub09,CleGar16}.

In this study we consider the additional source of tidal forces -- the tidal forces from the usual inflationary density perturbations in dark matter. This effect was not considered previously in the calculations of the PBHs pair evolution. These additional tidal forces alter the distribution of PBHs pairs over their orbital parameters, and finally suppress the rate of gravitational bursts by about half of the order of magnitude. 

The close PBHs pairs form at the cosmological stage of the radiation dominance (RD), but they can also form later 
at the stage of matter domination. A halo of dark matter is accumulated around the PBHs by the secondary accretion  mechanism  \cite{Ber85}. The PBHs in the halo  experience the dynamic friction, lose angular momentum and move toward the center, there they form 
the gravitationally bounded binary, even if they were not associated initially \cite{Hayetal09}. The further fate of this pair is uncertain, because the effectiveness of dynamic friction in this situation was not investigated.

The paper is organised as follows. In the Section~\ref{label} we mainly define the notations and repeat some calculations of  \cite{Naketal97,Sasetal16,Ioketal98} by other method. Section~\ref{label} is devoted to the new effect -- the tidal forces from adiabatic density perturbations with CMB-normalised spectrum. Then  in the Section~\ref{ratesec} we calculate the resultant rate of the gravitational bursts. In the Section~\ref{dmsec} we consider critically the possibility of the PBHs merge inside dark matter minihalos. And finally we present some conclusions in the Section~\ref{conclsec}.

%%%%%%%%%%%%

\section{The evolution of binaries at the stage of radiation domination}
\label{label}

Consider the evolution of the binary PBHs at the RD-stage, $t<t_{\rm eq}$, where $t_{\rm eq}$ is the moment of matter-radiation equality. Let the 
scale factor of the universe $a(t)$ is normalized as $a(t_{\rm eq})=1$, then the density of radiation  $\rho_r=\rho_{\rm eq}/a^4$.
Let us assume that the PBHs constitute the $f\leq1$ fraction of dark matter, $\Omega_{\rm BH}=f\Omega_m$. In the case $f<1$ the rest of dark matter can be in some other form, for example in the form of weakly interacting massive particles (WIMPS). Denote by $x$ the comoving space between the components of PBHs pair, while the
average distance between PBHs is
\begin{equation}
\bar{x}=n^{-1/3}=\left(\frac{M_{\rm BH}}{f\rho_{\rm eq}}\right)^
{1/3}.
\end{equation}
Let the binary form at  $t<t_{\rm eq}$. The condition for this is \cite{Naketal97,Sasetal16}
\begin{equation}
\frac{M_{\rm BH}}{x^3a_m^3}\sim\rho_r,
\end{equation}
where
\begin{equation}
a_m=\frac{1}{f}\left(\frac{x}{\bar{x}}\right)^3.
\label{amexpr}
\end{equation}
This occurs at the RD-stage if $a_m<1$ that 
can be written as $x<x^*$, where
\begin{equation}
x^*=\left(\frac{M_{\rm BH}}{\rho_{\rm eq}}\right)^{1/3}.
\end{equation}
The $x^*$ would be the mean distance between the PBH in the case $f=1$.

According to \cite{Naketal97,Sasetal16} the semi-major axis of 
orbit is fixed at the time given by (\ref{amexpr}), and the minor axis is influenced by tidal forces from the 
third PBH, located at the comoving distance $y<x$. 
Minor axis is calculated as the product of the tidal force and the 
time of free fall. As a result, \cite{Sasetal16} obtained for the major and minor semi-axes, respectively,
\begin{equation}
A=\alpha\frac{1}{f}\frac{x^4}{\bar{x}^3},\qquad B=\beta A\left(\frac{x}{y}\right)^3,
\label{ab}
\end{equation}
and the eccentricity of the orbit
\begin{equation}
e=\sqrt{1-\frac{B^2}{A^2}}=\sqrt{1-\beta^2\frac{x^6}{y^6}}.
\label{eexpr}
\end{equation}
In the paper \cite{Sasetal16} the expressions (\ref{ab}) and (\ref{eexpr}) 
were obtained in the case $\alpha=\beta=1$.  The \cite{Ioketal98} introduced the additional factors $\alpha$ and $\beta$, and obtained them from the numerical solution of the evolution equations. In the rest of this section we solve the similar equations and in the next section we consider additional effects. 

The distance of each PBH from the center of masses of the pair is $r/2$. We consider the scales under the 
cosmological horizon $r\ll ct$, so we can use the Newtonian 
dynamics, taking into account the contribution of uniformly 
distributed radiation by the substitution $\rho\to\rho+3pc^2$ \cite{McCrea51}. The $r$ evolves according to the equation
\begin{equation}
\frac{d^2(r/2)}{dt^2}=-\frac{GM_{\rm BH}}{r^2}-\frac{8\pi G  \rho_r (r/2)}{3},
\label{d2rdt1}
\end{equation}
where the last term describes the gravitational effect of uniformly distributed radiation within the sphere with radius $r/2$.
Following the approach of \cite{KolTka94}, we do the following parametrization
\begin{equation}
r=abx.
\label{abxi}
\end{equation}
The evolution of $b$ shows the factor of displacement from the comoving distance in the homogeneous universe.
If one takes into account that at the radiation-dominated stage $a\simeq t^{1/2}/t_{\rm eq}^{1/2}$, the Eq.~(\ref{d2rdt1}) can be rewritten as
\begin{equation}
a\frac{d^2b}{da^2}+\frac{db}{da}=-\frac{3}{4\pi b^2}\frac{x^{*3}}{x^3}.
\label{bigeq}
\end{equation}
Analogous equation was obtained in \cite{Sasetal16} by other method. The combination of the quantities in the right-hand side
\begin{equation}
\delta_i=\frac{2M_{\rm BH}}{(4\pi/3)x^3\rho_{\rm eq}}=\frac{3}{2\pi}\frac{x^{*3}}{x^3}
\label{deli}
\end{equation}
is the density perturbation produced by two PBH, so one can conclude that 
Eq.~(\ref{bigeq}) is equivalent to the equation of \cite{KolTka94} (taken in the limit $a\ll1$), which describes the 
evolution of spherically symmetric entropy (isocurvature) perturbations. Initial 
time $t_i$ is taken when the PBH mass is negligible in comparison with the radiation mass inside the sphere, and the initial 
conditions at the time $t_i$ have the form $b=1$, $db/da=0$. The distance between the PBHs stops growing at the time $t_s$, when $dr/dt=0$, which corresponds to the condition $db/da=-b/a$ \cite{KolTka94}. It is easy to see that  $\alpha=b(a(t_s))$ in the Eq.~(\ref{ab}).

Let us consider the evolution of small semi-axis under the influence of tidal forces from the 3rd PBH, located at the 
comoving distance $y$ from the pair. We consider the
transverse displacement $\delta r_t$ as a small perturbation to the radial orbit, and we introduce the
parametrization
\begin{equation}
\delta r_t=a\xi.
\label{xiparam}
\end{equation}
The tidal force has the form
\begin{equation}
F_t=\frac{2GM_{\rm BH}^2}{(ay)^3}r,
\label{ftid}
\end{equation}
and the equation $d^2\delta r_t/dt^2=F_t$ can be written 
as
\begin{equation}
a\frac{d^2\xi}{da^2}+\frac{d\xi}{da}-\frac{\xi}{a}=-\frac{3}{4\pi}\frac{x^{*3}xb}{y^3}.
\label{bigeqxi}
\end{equation}
The initial conditions at the time $t_i$ are $\xi=d\xi/da=0$.

\begin{figure}[t]
\begin{center}
\includegraphics[angle=0,width=0.45\textwidth]{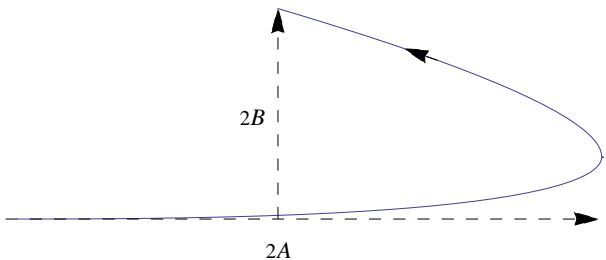}
\end{center}
\caption{Example of the orbit evolution under the influence of tidal forces  from the numerical solution of 
Eqs.~(\ref{bigeq}) and (\ref{bigeqxi}). The vertical and 
horizontal dashed lines show the doubled major and minor semi-axis of the 
orbit, respectively.} \label{grorbit}
\end{figure}

We solve the system of equations (\ref{bigeq}), (\ref{bigeqxi}) 
numerically until the binary expands up to the maximum distance (this 
distance is the $A$ major semi-axis) and then 
approaches twice in radius. The value of $a\xi$ at this moment gives the 
size of the minor axis $B$. We assume for simplicity that all PBH have 
the same masses $M_{\rm BH}=30M_\odot$. An example of the orbit
is shown at	Fig.~\ref{grorbit}.
Solving the equation (\ref{bigeq}) and (\ref{bigeqxi}) for any $x$ 
(different values of $y$ are taken into account in (\ref{bigeqxi}) by the scaling of $\xi$) and comparing with (\ref{ab}), we obtain the following correction factors 
in (\ref{ab})
\begin{equation}
\alpha\simeq0.64,\qquad \beta\simeq2.8.
\label{abcorr}
\end{equation}
In our calculations these factors are independent of $x$ with high accuracy, however \cite{Ioketal98} obtained $\alpha\simeq0.4$, $\beta\simeq0.8$ and weak dependence on the fraction $x/y$.
We don't consider here the angle dependence (direction to the 3rd PBH). This dependence was considered in \cite{Ioketal98}.

%%%%%%%%%%%%

\section{Tidal forces from inflationary perturbations}
\label{tidsec}

Neighbourhood PBHs are not the only sources of disturbing. Tidal forces can be produced also by the dark matter density perturbations,  and the main contribution comes from the  
characteristic scale $x^*$, where dark matter mass is equal to the PBHs mass $M\simeq M_{\rm BH}$. 
Really, the scale of the PBHs binary contains the mass of dark matter
\begin{equation}
M\simeq\rho_{\rm eq}x^3=M_{\rm BH}\frac{x^3}{x^{*3}}. 
\label{mdm}
\end{equation}
If $M<M_{\rm BH}$ the PBHs pair forms at the 
RD-stage, and from (\ref{mdm}) it follows $x<x^*$. Otherwise, the PBHs do not define the dynamics of the 
system, but they follow the motion of dark matter. At smaller scales $x<x^*$ the dark matter is highly disturbed by the PBHs and is involved into the motion. The tide from the smaller scales is unclear. In contrast, the larger scales produce regular tidal force, so we consider the effect of the scales $x\geq x^*$ as minimum warranted contribution.   Let the density perturbation at the mass scale $M$ at the moment $t_{\rm eq}$ is $\delta_{\rm eq}$.
R.m.s. value of perturbation is (see e.g. \cite{BerDokEro13} and references therein)
\begin{eqnarray}
\sigma_{\rm eq}(M)&\simeq &8.2\times 10^{3.7(n_s-1)-3}
 \left(\frac{M}{M_{\odot}}\right)^{\frac{1-n_s}{6}}
 \nonumber
\\
&\times & \left[1-0.06\log\left(\frac{M}{M_{\odot}}\right)\right]^{\frac{3}{2}}, \label{A5}
\end{eqnarray}
where $n_s=0.9608\pm0.0054$ according to the Planck data. In particular, $\sigma_{\rm eq}(30M_\odot)\simeq5.2\times10^{-3}$. At the RD-stage the adiabatic perturbation $\delta$ evolves only logarithmically. We can write the evolution of the dark matter perturbation as follows 
\begin{equation}
\delta=\delta_{\rm eq}F(a),
\end{equation} 
where the function \cite{BerDokEro13}
\begin{equation}
F(a)=\frac{\ln\left[g(a)/\sqrt{3}\right]+{\bf C}-\frac{1}{2}}
{\ln\left[g(1)/\sqrt{3}\right]+{\bf C}-\frac{1}{2}},
\end{equation}
${\bf C}\simeq0.577$ is the Euler constant, and 
\begin{equation}
g(a)=\frac{\pi}{2^{2/3}}\left(\frac{3}{2\pi}\right)^{1/6}
\frac{ac}{M^{1/3}G^{1/2}\rho_{\mathrm{eq}}^{1/6}}.
\label{xy}
\end{equation}

Then the dark matter excess can be estimated as $\Delta M\simeq M\delta$,
and the tidal force acting on the pair is
\begin{equation}
F_\sigma\simeq\frac{GM_{\rm BH}\Delta M (axb)}{(ax^*)^3},
\label{fsigma}
\end{equation}
where $b$ is defined by (\ref{abxi}). 
The comparison 
\begin{equation}
\frac{F_\sigma}{F_t}=\frac{\delta_{\rm eq}}{f}F(a)\frac{y^3}{\bar{x}^{3}}
\end{equation}
shows that for the sufficiently isolated PBHs pairs with $y\sim\bar{x}$ the contribution from inflation perturbations can prevail over the contribution from the 3rd PBH. The additional source of the tidal forces changes the distribution over eccentricity $e$, calculated in \cite{Naketal97,Sasetal16}. 

We substitute the (\ref{fsigma}) into the equation $d^2\delta r/dt^2=F_\sigma$ and solve this equation (instead of (\ref{bigeqxi})) numerically for different $x$. If we parametrise 
\begin{equation}
B=kA\frac{x^3}{x^{*3}}\nu,
\label{bbigk}
\end{equation}
where $\nu=\delta_{\rm eq}/\sigma_{\rm eq}$,
we obtain from the numerical solution that the factor $k$ is approximately constant, 
\begin{equation}
k\simeq0.005.
\end{equation}
In the next Section we use these results for the calculation of the rate of gravitational bursts.

%%%%%%%%%%%%

\section{Rate of gravitational bursts}
\label{ratesec}

Let us first neglect the effect of tidal forces from dark matter inflationary perturbations, but take into account the 
above correction factors (\ref{abcorr}). The probability distribution for $A$ and $e$, 
obtained in \cite{Naketal97} and \cite{Sasetal16}, can be rewritten in this case as
\begin{equation}
dP=\frac{3}{2}\frac{f^{3/2}\beta}{\alpha^{3/2}\bar{x}^{3/2}}A^{1/2}e(1-e^2)^{-3/2}dAde.
\label{dpdis}
\end{equation}
This distribution is obtained by calculating the Jacobian $\partial(x,y)/\partial(A,e)$ under the assumption of the uniform 
distribution of $x$ and $y<x$ in the ranges from $0$ to $\bar{x}$.
Lifetime of PBHs pair due to the radiation of gravitational waves is
\cite{Sasetal16}
\begin{equation}
t_c=\frac{5c^5}{512G^3M_{\rm BH}^3}A^4(1-e^2)^{7/2}.
\label{tc}
\end{equation}
The probability that the time (\ref{tc}) is less than $t$ can be 
obtained from (\ref{dpdis}) by integrating over the corresponding 
curved region in the parameter space of $A$ and $e$, selected by the condition $t_c<t$:
\begin{equation}
P(<t)=\frac{\alpha}{\beta}\left[\frac{37}{29}\left(\frac{t}{t_{\rm max}}\right)^{3/37}-\frac{8}{29}\left(\frac{t}{t_{\rm max}}\right)^{3/8}\right],
\label{pint}
\end{equation}
where
\begin{equation}
t_{\rm max}=\frac{5c^5}{512G^3M_{\rm BH}^3}\frac{\alpha^4}{\beta^{16/3}}\frac{x^{*4}}{f^{16/3}}.
\label{tmax}
\end{equation}
In the case $\alpha=\beta=1$ the distributions (\ref{dpdis}) and 
(\ref{pint}) coincide with those obtained in \cite{Naketal97}.

\begin{figure}[t]
\begin{center}
\includegraphics[angle=0,width=0.49\textwidth]{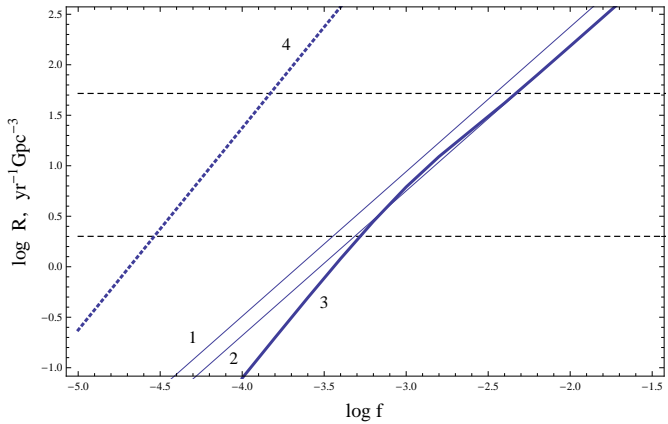}
\end{center}
\caption{The rate of gravitational waves bursts from the collision 
of PBHs in pairs, depending on the fraction $f$ of PBHs in dark matter. The upper solid curve 1 shows the result of the work \cite{Sasetal16}. The middle solid curve 2 was obtained according to (\ref{rate1}) with the correction factors (\ref{abcorr}). The lower bold curve 
3 shows the result of the current work according to (\ref{rate1}) and (\ref{dpdxdydnu}). Dotted curve 4 shows 
the rate of bursts under the assumption of the efficiency of dynamic 
friction inside minihalos at the last stage of PBHs pair evolution. 
The horizontal dashed lines limit the area of LIGO observations, obtained in 
\cite{Sasetal16}.} \label{grrate}
\end{figure}
The rate of gravitational wave bursts near the current moment $t_0$ is
\begin{equation}
R=\left.\frac{\rho_c\Omega_mf}{M_{\rm BH}}\frac{dP(<t)}{dt}\right|_{t=t_0},
\label{rate1}
\end{equation}
where $\rho_c=9.3\times10^{-30}$~g~cm$^{-3}$ is the 
critical density, $\Omega_m\approx0.27$. The result of the 
calculation is shown at Fig.~\ref{grrate} in comparison with the 
result of \cite{Sasetal16}, which is reproduced with $\alpha=\beta=1$.
Our rate  is about 40\% of \cite{Sasetal16} value.

Now let us take into account the effect of tidal forces from ordinary inflationary perturbations as it was considered in the Section~\ref{tidsec}. 
In this case we have the 3-dimensional probability distribution
\begin{equation}
dP=\frac{18x^2y^2}{\bar{x}^6}\frac{1}{\sqrt{2\pi}}e^{-\nu^2/2}dxdyd\nu,
\label{dpdxdydnu}
\end{equation}
where we consider not only positive $\nu>0$  but also negative $\nu<0$  perturbations. Now we have the sum of (\ref{ab}) and (\ref{bbigk})
\begin{equation}
B=\beta A\left(\frac{x}{y}\right)^3+kA\frac{x^3}{x^{*3}}\nu.
\label{bplus}
\end{equation}
The tidal force from inflation perturbations prevails if 
\begin{equation}
\nu>\nu^*\equiv\frac{\beta x^{*3}}{y^3k}.
\end{equation}
The integration of  (\ref{dpdxdydnu}) in the 3D space with the condition $t_c<t$  is more complicated in this case. The result of the numerical integration is shown at  Fig.~\ref{grrate}  by the curve 3. One can see that in the region $f\sim5\times10^{-4}-5\times10^{-3}$  the tidal forces from dark matter do not change the result significantly, but at smaller $f$  they can suppress the rate of the gravitational bursts by the factor  $\sim1.5-2$.

%%%%%%%%%%%%

\section{Binaries inside dark matter minihalos}
\label{dmsec}

Consider the case $f\ll1$ and $\delta_i<1$ in 
(\ref{deli}). The first of these conditions means that PBHs constitute only small fraction of dark matter and the rest of the 
dark matter consists of something else, for example, of new elementary 
particles. The second condition means that we consider sufficiently large region where PBHs
create only small perturbation in dark matter. This perturbation at $t>t_{\rm eq}$ evolves into the minihalo of dark matter around 
the PBHs pair. Similar model was proposed in \cite{Hayetal09}. Consider the evolution of perturbations at the comoving scale $x$ (the final minihalo contains the mass $M\sim10^2M_{\rm BH}$ \cite{BerDokEro13}). After the detachment  from the cosmological 
expansion the perturbed region is compressed twice and virialized at the final size \cite{Pee80}
\begin{equation}
r_v=\frac{x}{2\delta_i}.
\end{equation} 
The mass within this radius is given by (\ref{mdm}), so the velocity 
dispersion
\begin{equation}
v\simeq\left(\frac{GM}{r_v}\right)^{1/2}\simeq\left(\frac{3GM_{\rm BH}}{\pi x}\right)^{1/2},
\end{equation} 
and the average density
\begin{equation}
\rho_H\simeq\frac{3^4}{4\pi^4}\frac{x^{*9}}{x^9}\rho_{\rm eq}.
\end{equation}
PBHs experience dynamical friction, lose the orbital
angular momenta and approach the minihalos centers. The change of the orbital radius obeys
the equation \cite{Sasl}
\begin{equation}
\frac{dr}{dt}=-\frac{4\pi G^2M_{\rm BH}\rho_H(r)Dr}{v(r)^3},
 \label{difdyn}
\end{equation}
where numerical constant $D\approx4.27$. From Eq.~(\ref{difdyn}) we obtain the characteristic time of the fall to the center
\begin{equation}
t_d\simeq5.3t_{\rm eq}\left(\frac{x}{x^*}\right)^{15/2}.
\end{equation}
PBHs come to the radius $r$, within which the mass of dark matter is $M(r)\sim M_{\rm BH}$, and the PBHs sit 
on the orbit around each other. Dark matter within 
their orbit and the surrounding material will lead to the evolution 
of the orbit, but the effectiveness of dynamical friction under these 
conditions has not been elucidated yet, and it requires further 
investigation. The lifetime of the pair (\ref{tc}) is many 
orders of magnitude greater than the age of the universe $t_0$, 
therefore, for the further approach and merger the PBHs must continue to lose the orbital angular momenta. 

Virial velocity in the minihalo is $v\simeq0.36$~km~s$^{-1}$, and the corresponding virial temperature of baryonic gas is $T\simeq m_pv^2/(2k_B)\sim8$~K, where $k_B$ is the Boltzmann constant. So the radiative cooling mechanisms do not work, and the gas does not 
form stars, which could give additional dynamical friction.

Consider the most optimistic case, when at the radii $<r$ the dynamic 
friction is as effective as before and is described 
by (\ref{difdyn}). Then the significant fraction of the PBHs pairs can
merge during the Hubble time and produce gravitational 
bursts. Distribution of $x$ under the assumption of uniformity \cite{Naketal97,Sasetal16} has the simple form
\begin{equation}
dP=\frac{3x^2dx}{\bar{x}^3}.
\end{equation}
The condition $t_d<t$ selects the region of integration, and
\begin{equation}
P(<t)\simeq64f\left(\frac{t}{t_0}\right)^{2/5}.
\end{equation}
Substituting into (\ref{rate1}), we obtain the rate of gravitational bursts that is shown at Fig.~\ref{grrate} by the point curve. This result was obtained under the assumption of high efficiency of dynamical friction at the last stage of evolution, so it gives the most optimistic scenario, but this model requires 
further elaboration. 

%%%%%%%%%%%%%%%%

\section{Conclusion}
\label{conclsec}

In this work we considered the formation and evolution of the PBHs pairs and found the expected rate of LIGO 
gravitational bursts. Our approach is similar to one of \cite{Naketal97,Ioketal98,Sasetal16}, but the 
calculations are made by another method, which takes into account the tidal forces from adiabatic density perturbations of dark matter with CMB-normalised spectrum. These tidal forces alter the probability distribution of PBHs pairs over their orbital parameters, and suppress the rate of the gravitational bursts can suppress the rate of the gravitational bursts by the factor $\sim1.5-2$. It turned out that the merge rate matches the LIGO signals if the PBHs constitute the fraction $f\sim5\times10^{-4}-5\times10^{-3}$ of dark matter. The rate of the bursts could be higher due to the formation of dark matter minihalos around PBHs \cite{Hayetal09}, but this scenario is questionable because of the uncertain efficiency of dynamical friction at the last stage of pair evolution.

%%%%%%%%%%%%

\end{document}